\documentclass[11pt,aps,prx,tightenlines,onecolumn,floatfix,superscriptaddress,longbibliography]{revtex4-2}
\usepackage{graphicx}
\usepackage{float}
\usepackage{amsfonts}
\usepackage{amsmath}
\usepackage{MnSymbol}
\usepackage{xcolor}

\begin{document}
\title{Real-time operator evolution in two and three dimensions via sparse Pauli dynamics}
\author{Tomislav Begu\v{s}i\'{c}}
\author{Garnet Kin-Lic Chan}
\email{gkc1000@gmail.com}
\affiliation{Division of Chemistry and Chemical Engineering, California Institute of Technology, Pasadena, California 91125, USA}
\date{\today}

\begin{abstract}
We study real-time operator evolution using sparse Pauli dynamics, a recently developed method for simulating expectation values of quantum circuits. On the examples of energy and charge diffusion in 1D spin chains and sudden quench dynamics in the 2D transverse-field Ising model, it is shown that this approach can compete with state-of-the-art tensor network methods. We further demonstrate the flexibility of the approach by studying quench dynamics in the 3D transverse-field Ising model which is highly challenging for tensor network methods.
For the simulation of expectation value dynamics starting in a computational basis state, we introduce an extension of sparse Pauli dynamics that truncates the growing sum of Pauli operators by discarding terms with a large number of $X$ and $Y$ matrices. This is validated by our 2D and 3D simulations. Finally, we argue that sparse Pauli dynamics is not only capable of converging challenging observables to high accuracy, but can also serve as a reliable approximate approach even when given only limited computational resources.
\end{abstract}

\maketitle

\graphicspath{{./figures/}}

\section{Introduction}
\label{sec:intro}
Numerical simulations of quantum dynamics are essential for our understanding of strongly correlated many-body physics. For large systems and long-time dynamics, exact state-vector or tensor network contraction~\cite{gray2021hyper,huang2021efficient,gray2022hyper} methods become computationally intractable and approximate numerical methods are required. These can be formulated in different pictures, for example, the Schr\"odinger picture, where the state is evolved, or the Heisenberg picture, where the state is unchanged and time evolution is applied to the observable.

Operator time evolution appears naturally in the dynamics of high-temperature systems \cite{rakovszky2022dissipation} and the theory of operator spreading \cite{Nahum2018, VonKeyserlingk2018, Khemani2018}, and has proven useful in the computation of time-correlation functions \cite{VonKeyserlingk2022} and out-of-time-order correlations \cite{Mi2021}. Compared to the Schr\"odinger picture, working in the Heisenberg picture has the benefit that one can take advantage of the dynamical light-cone structure, i.e., the fact that the support of some relevant observables is initially local and grows in time. Operator time evolution has been the subject of tensor network studies, e.g., those based on matrix-product operators (MPO) \cite{Verstraete2004, Zwolak2004, rakovszky2022dissipation, Anand2023} or projected entangled-pair operators (PEPO) \cite{Alhambra2021,liao2023simulation}, whose performance depends on the degree of operator entanglement in the evolved observable.

Alternative Heisenberg-picture methods have been formulated to take advantage of the sparsity of the observable in the Pauli operator representation. 
Under time evolution, local observables spread to a superposition of an increasing number of Pauli operators, and 
various strategies to curb the (at worst) exponential growth of the number of Pauli operators have been employed, including stochastic sampling of Pauli paths \cite{rall2019simulation} or discarding Pauli paths based on Pauli weights \cite{Shao2023}, perturbation order with respect to the nearest Clifford transformation \cite{begusic2023simulating}, Fourier order \cite{Nemkov_Fedorov:2023}, or combinations thereof \cite{fontana2023classical, rudolph2023classical}. These methods have mainly been developed in the context of simulating quantum circuit expectation values, where the number of Pauli operators can be low due to the use of Clifford gates or noise \cite{Aharonov2023, Schuster2024}.

Here, we consider one of these methods, namely the sparse Pauli dynamics (SPD) \cite{Begusic2024}, which was applied successfully recently to simulate the kicked transverse-field Ising model quantum simulation of Ref.~\cite{kim2023evidence}. There, we showed that it can simulate the expectation values faster than the quantum device to reach comparable accuracy and that, given more time, can produce more accurate results than the quantum processor. These simulations also demonstrated that SPD performs well compared to other classical simulation strategies based on state \cite{Kechedzhi2024, Tindall2024, Patra2024} and operator evolution \cite{Anand2023, rudolph2023classical, Shao2023, liao2023simulation}. In this work, we analyze its performance in simulating real-time dynamics on 1D spin chains, where we compare it to MPO dynamics including its recent dissipation assisted variant \cite{rakovszky2022dissipation}, as well as in the transverse-field Ising model on square (2D) and cubic (3D) lattices, where we use available 2D tensor network simulation benchmarks. For the latter, we introduce a modification of the original method that further reduces the computational cost of simulating time-dependent expectation values when the initial state is a computational basis state (e.g., $| 0^{\otimes n} \rangle$).

\section{Real-time sparse Pauli dynamics}
\label{sec:spd}
In SPD we write the observable operator
\begin{equation}
    O = \sum_{P \in \mathcal{P}} a_P P
    \label{eq:O_def}
\end{equation}
as a sum of Pauli operators $P$ with complex coefficients $a_P$, where $\mathcal{P}$ is a set of Pauli operators that contribute to $O$. The central part of our algorithm is the action of a Pauli rotation operator $U_{\sigma}(\theta) = \exp(-i \theta \sigma / 2)$ on the observable (\ref{eq:O_def}):
\begin{equation}
    \tilde{O} = U_{\sigma}(\theta)^{\dagger} O U_{\sigma}(\theta) = \sum_{P \in \mathcal{P}_C} a_P P + \sum_{P \in \mathcal{P}_A} (a_P \cos(\theta) P + i a_P \sin(\theta) \sigma \cdot P),
    \label{eq:O_rot}
\end{equation}
which follows from
\begin{equation}
    U_{\sigma}(\theta)^{\dagger} P U_{\sigma}(\theta) = \begin{cases}
        P, \, \, \qquad \qquad \qquad \qquad \qquad [P, \sigma] = 0, \\
        \cos(\theta) P + i \sin(\theta) \sigma \cdot P, \quad \{P, \sigma\} = 0,
    \end{cases}
    \label{eq:P_rot}
\end{equation}
where $[\cdot, \cdot]$ denotes a commutator and $\{\cdot, \cdot\}$ an anticommutator. In Eq.~(\ref{eq:O_rot}), $\mathcal{P}_C$ ($\mathcal{P}_A$) is a set of Pauli operators in $\mathcal{P}$ that commute (anticommute) with $\sigma$. The right-hand side of Eq.~(\ref{eq:O_rot}) can be brought into the form of Eq.~(\ref{eq:O_def}) by identifying which $\sigma \cdot P$ already exist in $\mathcal{P}$ and which have to be added to represent the rotated observable $\tilde{O}$. In general, the number of Pauli operators that we have to store, $N = |\mathcal{P}|$, grows exponentially with the number of unitary Pauli rotation operators applied. To limit the growth of $N$, we replace the exactly rotated observable $\tilde{O}$ by
\begin{equation}
    \tilde{O}_{\delta} = \Pi_{\delta} (\tilde{O}), 
\end{equation}
where $\Pi_{\delta}$ acts by discarding all Pauli operators $P$ with $|a_P| < \delta$. Here, the threshold $\delta$ defines the approximation error, i.e., $\delta = 0$ corresponds to exact dynamics. In practice, the goal is to converge the simulation with respect to this tunable parameter. Additional implementation details can be found in Appendix~\ref{app_sec:spd}.

For real-time dynamics under a Hamiltonian
\begin{equation}
    H = \sum_j c_j H_j,
\end{equation}
where $c_j$ are real coefficients and $H_j$ are Pauli operators, we replace the exact time-evolution operator $U_{\Delta t} = \exp(- i H \Delta t)$, corresponding to a time step $\Delta t$, by the first-order Trotter splitting formula
\begin{equation}
    U_{\Delta t} \approx \prod_j \exp(- i c_j \Delta t H_j).
    \label{eq:U_deltat}
\end{equation}
Now the real-time evolution takes the form of applying multiple Pauli rotation gates, which allows us to use the SPD method as defined above. The size of the time step determines not only the Trotter error in Eq.~(\ref{eq:U_deltat}) but also the truncation error in SPD. Specifically, for small values of threshold $\delta$ and time step $\Delta t$, the error is a function of their ratio $\delta / \Delta t$. To show this, let us consider one substep of the dynamics in which we apply one Pauli rotation operator $\exp(- i c_j \Delta t H_j)$ to the observable (\ref{eq:O_def}), the result of which is shown in Eq.~(\ref{eq:O_rot}) with $\theta = 2 c_j \Delta t$. For a sufficiently small time step, we can expand Eq.~(\ref{eq:O_rot}) up to first order in $\Delta t$,
\begin{equation}
    \tilde{O} \approx \sum_{P \in \mathcal{P}_C} a_P P + \sum_{P \in \mathcal{P}_A} (a_P  P + 2 i a_P c_j \Delta t \sigma \cdot P) = O + 2 i c_j \Delta t \sum_{P \in \mathcal{P}_A} a_P H_j \cdot P.
\end{equation}
For small threshold $\delta$, we can assume that the threshold-based truncation will mainly discard Pauli operators $H_j \cdot P$ that do not already exist in $O$. The condition under which they are discarded reads $|2 c_j \Delta t a_P| < \delta$, i.e., $|a_P| < (\delta / \Delta t) / 2|c_j|$. Consequently, the truncation error and the number of Pauli operators depend only on the ratio $\delta / \Delta t$. Since the number of Pauli operators determines the computational cost per time step, it is generally preferred to use a larger time step---at fixed $\delta / \Delta t$, the cost per time step is constant, but a larger time step requires less steps to reach a given total time. In turn, the size of the time step is limited by the Trotter error, which must be validated before converging the calculation with respect to truncation threshold $\delta$.

Finally, we introduce an upper bound on the number of Pauli operators ($N$) for a given threshold $\delta$. Let us assume that the scaled Frobenius norm squared of the observable operator $O_0$ before propagation is
\begin{equation}
    \frac{1}{2^n} \text{Tr}(O_0^{\dagger}O_0) = c_0.
\end{equation}
Then, during propagation, $O_t = \sum_P a_P P$, and this norm is preserved under unitary dynamics but is otherwise smaller than $c_0$ if we truncate the observable based on sparse Pauli dynamics. Therefore, we have
\begin{equation}
    \frac{1}{2^n} \text{Tr}(O_t^{\dagger}O_t) = \sum_P |a_P|^2 \leq c_0.
\end{equation}
Because of the truncation based on the threshold $\delta$, we know that
\begin{equation}
    \sum_P |a_P|^2 \geq \sum_P \delta^2 = N \delta^2,
\end{equation}
which leads to
\begin{equation}
    N \leq \frac{c_0}{\delta^2}.
\end{equation}

\section{Related work and theoretical guarantees}

Sparse Pauli dynamics was developed in the context of quantum circuit simulation and originates from the idea that the sparse Pauli representation can be efficiently updated under Clifford operations \cite{Begusic2024, begusic2023simulating, Nemkov_Fedorov:2023}. In turn, for a fixed threshold, it typically exhibits an exponential scaling of the number of Pauli operators with respect to the number of non-Clifford gates. One way to analyze methods based on the sparse Pauli representation is by considering magic monotones, such as the recently developed operator stabilizer entropy \cite{Dowling_Turkeshi:2024}. This magic monotone, which has been shown to satisfy the Lieb-Robinson bound \cite{Lieb_Robinson:1972}, is closely related to the number of Pauli operators one must store to retain a faithful representation of the observable operator.

\begin{figure}
    \centering
    \includegraphics[width=0.55\linewidth]{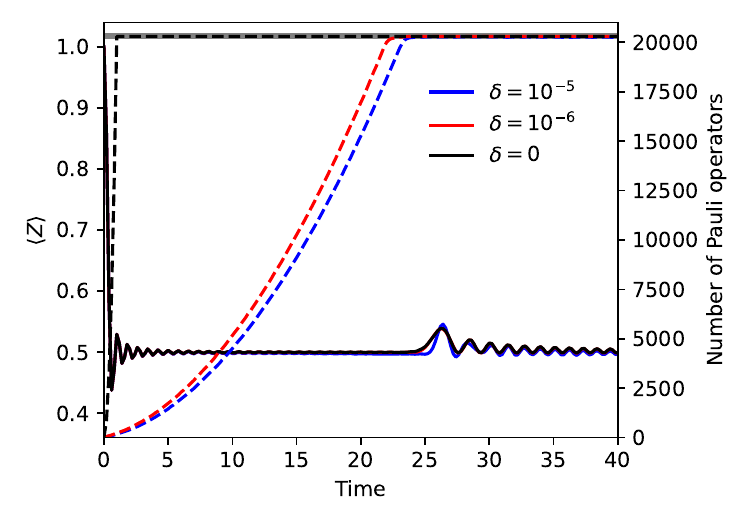}
    \caption{An example of observable dynamics in an integrable system for which the observable is limited to a polynomial number of Pauli operators in system size $n$. We compute $\langle Z_0 \rangle$, where the zero index indicates the central site of a length $n=101$ chain, time evolved with the transverse-field Ising model $H = \sum_i X_i X_{i+1} + Z_i$, starting from an all-zero state $|0^{\otimes n}\rangle$. The dynamics is simulated for $t=40$ with $\Delta t = 0.02$ and different values of the threshold $\delta$. Solid lines correspond to the expectation values (left axis), while the dashed lines correspond to the number of Pauli operators at each time step (right axis). The horizontal gray line is the maximum number of Pauli operators $\binom{2n}{2} = 20301$ that can be generated in this case.}
    \label{fig:integrable}
\end{figure}

One limit in which sparse Pauli dynamics is guaranteed to perform well is when the dynamics is constrained to a subset of Pauli operators that is much smaller than the operator Hilbert space (e.g., of polynomial size in $n$). This happens for certain observable evolutions in integrable systems, such as in the non-interacting transverse-field Ising chain. In Fig.~\ref{fig:integrable}, we show an example of such dynamics where the number of Pauli operators is limited. For the studied example, the Pauli operator $Z$ maps to a two-index Majorana operator and the Hamiltonian maps to a free fermion model, implying that the number of Paulis in the sparse Pauli representation saturates at $\binom{2n}{2}$ \cite{Barouch1971, Essler2016}. Finite-threshold sparse Pauli dynamics still provides a significant computational advantage over exact simulation due to the fact that many Paulis do not contribute to the observable at early times (compare the $\delta = 0$ and $\delta \neq 0$ results in Fig.~\ref{fig:integrable}). The above mentioned operator stabilizer entropy has also been derived for the dual-unitary XXZ model, a type of local interacting integrable circuit, where it was shown to saturate to a constant at long times \cite{Dowling_Turkeshi:2024}, thus providing another analytical example where the sparse Pauli representation can be shown to be asymptotically efficient.

Finally, Pauli-based techniques have been of central interest for analyzing the classical simulability of noisy quantum circuits, including the computation of probability amplitudes and expectation values of Pauli operators. Specifically, it has been shown that they enable simulation of noisy quantum circuits with a cost that is exponential only in the inverse of the noise rate but polynomial in the size of the system and depth of the circuit \cite{Aharonov2023, Shao2023, rudolph2023classical, Schuster2024}. More recently, these results have been generalized to noiseless, locally scrambling circuits \cite{Angrisani_Huang:2024}. Sparse Pauli dynamics profits directly from these theoretical results even though the theoretical analysis is based primarily on weight-based Pauli truncation, as there is a strong correlation between high weight Paulis and small coefficients.

However, unlike the above cases that come with some theoretical guarantees, general quantum dynamics comes with no guarantees and is expected to require computational resources that scale exponentially in time. This is shared by all classical numerical methods for simulating quantum dynamics. Therefore, in this setting we can analyze the merits of individual methods only empirically. In remainder of this paper we therefore focus on an empirical analysis of the numerical results we obtain for real-time quantum dynamics.

\section{Results}
\label{sec:res}
\subsection{Spin and energy diffusion constants in 1D chains}
\label{subsec:1D}
We begin with the computation of diffusion constants of conserved densities in spin chains of length $L$ at infinite temperature. The diffusion constant \cite{Bertini2021}
\begin{equation}
    D = \frac{1}{2} \frac{\partial}{\partial t} d^2(t)
\end{equation}
is defined through the time-derivative of the mean-square displacement
\begin{equation}
    d^2(t) = \sum_j C_j(t) j^2 - \left( \sum_j C_j(t) j \right)^2,
    \label{eq:d2_t}
\end{equation}
where
\begin{equation}
    C_j(t) = \frac{\text{Tr}[q_j q_{\frac{L+1}{2}}(t)]}{\left\lbrace \sum_j \text{Tr}[q_j q_{\frac{L+1}{2}}(0)] \right\rbrace}
    \label{eq:C_j_t}
\end{equation}
are the dynamical correlations between the operators $q$ localized at the central site $(L+1)/2$ and sites $j$. These operators represent conserved densities in the sense that $\sum_j q_j(t) = \sum_j q_j(0)$, which leads to $\sum_j C_j(t) = \sum_j C_j(0) = 1$.

This problem is naturally formulated in the Heisenberg picture and we employ SPD to numerically evolve $q_{\frac{L+1}{2}}(t)$ in time. Within the sparse Pauli representation of the operator, it is also easy to evaluate the overlap with another Pauli operator (or a sum of Pauli operators) $q_j$ (see Appendix~\ref{app_sec:spd}). Since the conservation laws are not strictly obeyed by the non-unitary truncation scheme employed in SPD, we replace the denominator in Eq.~(\ref{eq:C_j_t}) by $\sum_j \text{Tr}[q_j q_{\frac{L+1}{2}}(t)]$, i.e., we renormalize the correlations at post-processing.

\subsubsection{Models}
\label{subsubsec:1D-models}
Below, we introduce the examples of Ref.~\cite{rakovszky2022dissipation}, which were studied using dissipation assisted operator evolution (DAOE) combined with a MPO representation of the operator. There, the authors introduced an artificial dissipator that reduces the entanglement of the said MPO and demonstrated that this computational strategy is well founded for the simulation of diffusive transport at high temperature.

The first example is the one-dimensional tilted-field Ising model \cite{Kim2013, rakovszky2022dissipation}
\begin{equation}
    H = \sum_{j=1}^{L-1} Z_{j} Z_{j+1} + \sum_{j=1}^{L} (1.4 X_j + 0.9045 Z_j)
    \label{eq:H_tfim}
\end{equation}
 with open boundary conditions, while the conserved densities are the local energies
 \begin{equation}
     q_j = \frac{1}{2} Z_{j-1} Z_j + \frac{1}{2} Z_j Z_{j+1} + 1.4 X_j + 0.9054 Z_j.
     \label{eq:q_tfim}
 \end{equation}
$L=51$ is the number of sites in the chain.

The second model is the XX-ladder Hamiltonian \cite{rakovszky2022dissipation, Steinigeweg2014, Karrasch2015, Kloss2018}
\begin{equation}
    H = \frac{1}{4} \sum_{j=1}^{L-1}\sum_{a=1,2} (X_{j,a} X_{j+1, a} + Y_{j,a} Y_{j+1,a}) + \frac{1}{4}\sum_{j=1}^{L} (X_{j,1} X_{j, 2} + Y_{j,1} Y_{j,2}),
    \label{eq:H_xxladder}
\end{equation}
where the total spin is a conserved property and we consider the diffusion of $q_j = (Z_{j,1} + Z_{j,2})/2$ along the chain of length $L=41$ (number of sites is $n=2L=82$).

For these two models, we simulated the mean-square displacement (\ref{eq:d2_t}) for a total time of $t = 20$, with a time step of $\Delta t = 0.02$, unless stated otherwise. The diffusion constant was computed by linear regression of $d^2(t)$ between $t=10$ and $t=20$.

\subsubsection{Numerical results}
\label{subsubsec:1D-res}
\begin{figure}[!pth]
    \centering
    \includegraphics[width=0.39\textwidth]{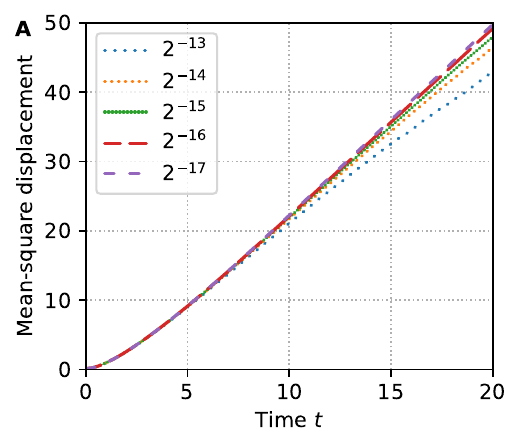}
    \includegraphics[width=0.4\textwidth]{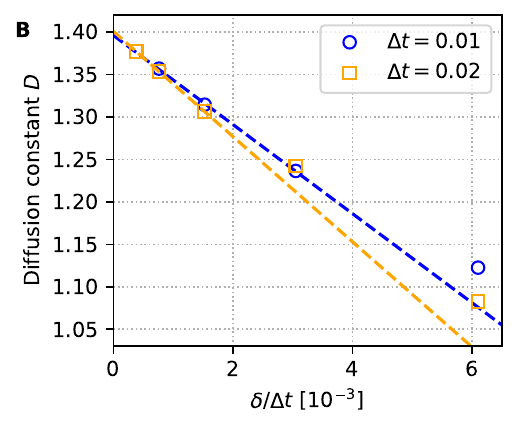}
    \includegraphics[width=0.39\textwidth]{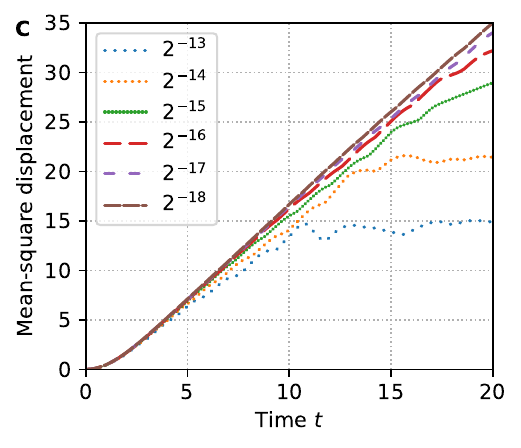}
    \includegraphics[width=0.4\textwidth]{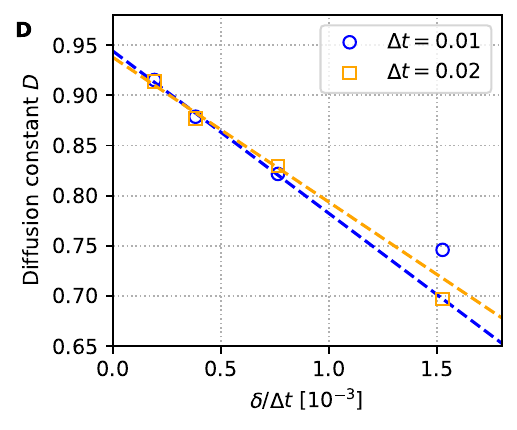}
    \caption{Time-dependent mean-square displacements (A, C) and diffusion constants (B, D) for the tilted-field Ising (A, B) and XX-ladder (C, D) models described in Sec.~\ref{subsubsec:1D-models}. Extrapolations correspond to linear fits to the leftmost three points.}
    \label{fig:1d}
\end{figure}

Figure~\ref{fig:1d} shows the mean-square displacements for the tilted-field Ising (panel A) and XX-ladder models (panel C), simulated using SPD with different values of the threshold $\delta$ ranging from $2^{-18}$ to $2^{-13}$. We can observe how $d^2(t)$ converges onto a straight line at large $t$ as we reduce the threshold and make the simulations more accurate. Figures~\ref{fig:1d}B, D plot the corresponding diffusion constant as a function $\delta/\Delta t$ for two sets of data, one using $\Delta t = 0.01$ and the other using $\Delta t = 0.02$. For this property, the results are amenable to an extrapolation in $\delta \to 0$, and we find $D \approx 1.4$ for the tilted-field Ising chain (same value was reported in Ref.~\cite{rakovszky2022dissipation}, whereas somewhat larger values of $D \approx 1.44$ \cite{YiThomas_White:2024} and $D \approx 1.55$ \cite{Artiaco_Badarson:2024} were found in more recent works), while $D \approx 0.94$ for the XX-ladder (compared to $D \approx 0.95$ of Refs.~\cite{Steinigeweg2014, Kloss2018, Wang_Gemmer:2024} and $D \approx 0.96-0.98$ reported in Ref.~\cite{rakovszky2022dissipation}). The plots of the diffusion constant also reveal the scaling relationship discussed in Sec.~\ref{sec:spd}. Namely, as the threshold is reduced, the simulations using different $\delta$ and $\Delta t$ but the same $\delta / \Delta t$ become closer to each other. In addition, we can numerically verify that using a larger time step reduces the computational cost. For example, the two points in Fig.~\ref{fig:1d}D at fixed $\delta / \Delta t = 2^{-18}/0.02 \approx 0.00019$ take around $84\,$h ($\Delta t = 0.01$, blue circle) and $43\,$h ($\Delta t = 0.02$, orange square) to simulate on 6 cores.

To illustrate that the above problems are non-trivial to simulate, we present a comparison between SPD and matrix-product operator (MPO) dynamics without dissipation assistance (Fig.~\ref{fig:1d-mpo}, see details in Appendix~\ref{app_sec:mpo}), a standard method for 1D dynamics.
Using a MPO bond dimension up to $\chi=2^9$ we find that the MPO simulations are still far from the exact results even for shorter chains ($L=9-21$) of the tilted-field Ising model, whereas SPD is more accurate already at rather large values of $\delta$. Similarly, Ref.~\cite{rakovszky2022dissipation} reported that the MPO dynamics of the $L=51$ chain with bond dimension $\chi = 2^9$ diverges from the exact result already at $t=8$. For the same system, SPD is visually well converged up to $t=15$ already at $\delta = 2^{-15}$.

\begin{figure}[!pth]
    \centering
    \includegraphics[width=\textwidth]{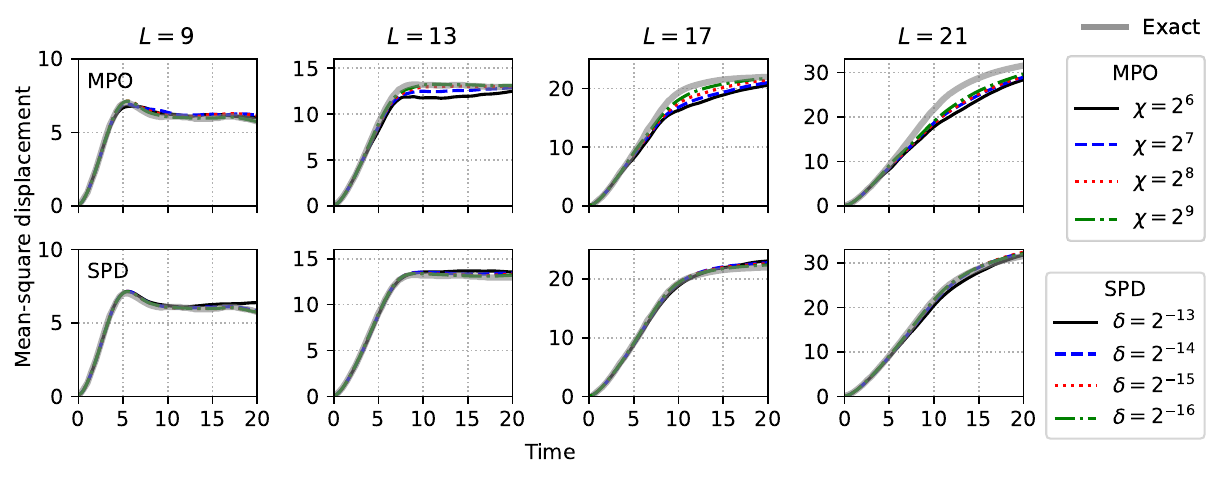}
    \caption{Comparison of SPD and MPO on the example of short-chain ($L=9-21$) tilted-field Ising model of Sec.~\ref{subsubsec:1D-models}. Exact benchmarks were adapted from Ref.~\cite{rakovszky2022dissipation}. Typical runtimes of MPO and SPD simulations are given in Table~\ref{tab:runtimes}, Appendix~\ref{app_sec:mpo}.}
    \label{fig:1d-mpo}
\end{figure}

\begin{figure}[!pth]
    \centering
    \includegraphics[width=0.435\textwidth]{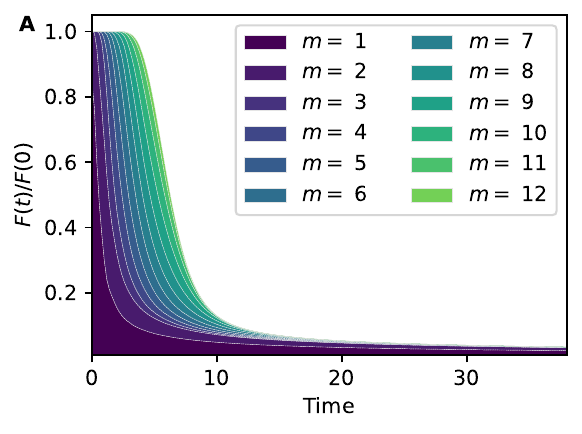}
    \includegraphics[width=0.45\textwidth]{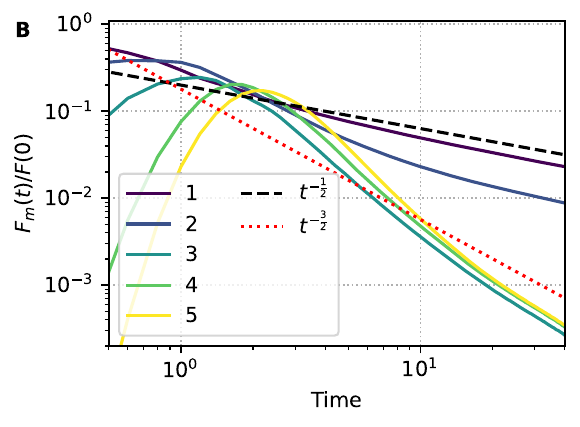}
    \caption{A: Contributions of Paulis of different weights $m$ to the total Frobenius norm squared $F$ of the operator $q_{\frac{L+1}{2}}(t)$ [Eq.~(\ref{eq:q_tfim})] evolved under the tilted field Ising Hamiltonian [Eq.~(\ref{eq:H_tfim})] (see Sec.~\ref{subsubsec:1D-models}). B: Individual Pauli weight contributions $F_m$ exhibit expected asymptotic behavior of $t^{-1/2}$ for conserved properties (local energy that consists of $m=1$ and $m=2$ Paulis) and $t^{-3/2}$ for other Pauli operators ($m>2$). In these examples, the simulations were performed up to $t=40$ with a threshold of $2^{-17}$. All data were rescaled by $F(0) = \text{Tr}[q_{\frac{L+1}{2}}(0)^2]/2^n = 3.27812025$.}
    \label{fig:1d-analysis}
\end{figure}

In the remainder of this Section, we analyze the operator evolution in terms of contributions from Pauli operators of different weights. Here, the Pauli weight is defined as the number of non-identity Pauli matrices in a Pauli operator. To this end, for any operator $O=\sum_{P \in \mathcal{P}} a_P P$, we introduce $F_m = \sum_{P \in \mathcal{P}_m} |a_P|^2$, where $\mathcal{P}_m \subseteq \mathcal{P}$ is a subset of Pauli operators with Pauli weight $m$. Then the sum $F = \sum_{m=1}^{n} F_m$ is constant for unitary dynamics and equal to the square of the Frobenius norm, i.e., $F = \text{Tr}(O^{\dagger}O)/2^n$. Figure~\ref{fig:1d-analysis}A shows the breakdown of $F$ into individual components $F_m$ (up to $m=12$) for the example of tilted-field Ising model [Eqs.~(\ref{eq:H_tfim}) and (\ref{eq:q_tfim})], demonstrating how the dynamics evolves initially low-weight Pauli operators into a sum of operators with higher weights. After some time, $F$ of the operator evolved with SPD deviates from its initial value because of threshold-based truncation. This truncation appears to affect high-weight Pauli operators more than the low-weight Paulis. This is confirmed in Fig.~\ref{fig:1d-analysis}B, where we show $F_m$ contributions for $m$ up to 5. As shown in previous works \cite{Khemani2018, Rakovszky2018, rakovszky2022dissipation}, Pauli operators corresponding to the local conserved quantity (here $m=1$ and $m=2$ for the local energy (\ref{eq:q_tfim})) obey the long-time scaling $F_m \sim t^{-1/2}$, whereas other Pauli operators (here $m>2$) obey $F_m \sim t^{-3/2}$ in the long-time limit.

This observation motivates methods that truncate the time-evolved operator based on Pauli weights, including the dissipation assisted operator evolution (DAOE) approach \cite{rakovszky2022dissipation}, low-weight simulation algorithm (LOWESA) \cite{fontana2023classical, rudolph2023classical} and observable's back-propagation on Pauli paths (OBPPP) \cite{Shao2023}, two approaches that were originally developed for the simulation of noisy quantum circuits, and the restricted state space approximation \cite{kuprov2007polynomially, Kuprov2008, Karabanov2011, Hogben2011}, introduced in the context of simulating nuclear magnetic resonance experiments. Our results indicate that SPD can take advantage of such behavior without explicitly truncating the sum based on Pauli weights because high-weight Pauli operators appear with small coefficients that do not meet the threshold criterion.

\subsection{Time-dependent expectation values}
\label{subsec:exp_val}
\subsubsection{X-truncated sparse Pauli dynamics}
\label{subsubsec:xspd}
In the examples above, we focused on infinite-temperature time-correlation functions of a time-evolved operator with a local, low-weight Pauli operator. Here, we consider expectation values of the form $\langle O \rangle_t = \langle 0^{\otimes n} | O(t) | 0^{\otimes n} \rangle$, where high-weight Pauli operators can contribute as long as they are composed only of identity and $Z$ matrices (i.e., if they are diagonal in the computational basis). Default Heisenberg evolution does not account for the information about the state over which we take the expectation value but rather treats all Pauli operators equally. Within the framework of SPD, this means that we keep a large number of Pauli operators, of which only a fraction contribute to the observable.

To further truncate the number of Pauli operators without introducing large errors in the expectation value $\langle O \rangle_t$, we propose to discard Pauli operators composed of more than $M$ $X$ or $Y$ Pauli matrices. We refer to the number of $X/Y$ matrices as the $X$-weight of a Pauli operator and we call this additional truncation scheme X-truncated SPD (xSPD). The truncation introduces the additional assumption that for certain (short) times, there is limited operator backflow from high $X$-weight Paulis to the manifold of $Z$-type Pauli operators. For each calculation, we test the value of $M$ to ensure that the error introduced by the X-truncation scheme is sufficiently small (for example, smaller than the target convergence criterion). The X-truncation is applied only every $T$ steps of the dynamics to limit the impact on the accuracy. In our calculations, we fixed $T$ to 5 steps. Finally, we note here that alternatives to our hard $M$ cutoff could also be considered, such as introducing an artificial dissipator based on the Pauli's X-weight that would be similar in spirit to DAOE \cite{rakovszky2022dissipation}.

In the following, we apply this modification of the original SPD method to dynamics in the 2D (square lattice) and 3D (cubic lattice) transverse-field Ising models described by the Hamiltonian
\begin{equation}
    H = -\sum_{\langle j k \rangle} X_{j} X_{k} - h \sum_{j=1}^{L}  Z_j,
    \label{eq:H_tfim_2}
\end{equation}
where the first sum runs over nearest neighbors on the lattice with open boundary conditions and $h$ controls the magnitude of the field. We consider the time-dependent magnetization $\langle Z\rangle_t = \langle 0^{\otimes n}| Z_0(t) |0^{\otimes n}\rangle$, where $Z_0$ denotes the $Z$ Pauli operator on the central site. Physically this corresponds to the magnetization induced after a sudden quench from infinite $h$ to a finite value of $h$.

In this setting, the first-order Trotter splitting
\begin{equation}
    U^{(1)}(t) = \left[e^{ i \Delta t \sum_{\langle j k \rangle} X_{j} X_{k}} e^{ i h \Delta t \sum_{j=1}^{L}  Z_j}\right]^K,
\end{equation}
where $K = t/\Delta t$ is the number of time steps, is equivalent to the second-order splitting
\begin{eqnarray}
    U^{(2)}(t) &=& \left[e^{ i \frac{1}{2} h \Delta t \sum_{j=1}^{L}  Z_j} e^{ i \Delta t \sum_{\langle j k \rangle} X_{j} X_{k}} e^{ i \frac{1}{2} h \Delta t \sum_{j=1}^{L}  Z_j}\right]^K \\ 
    &=& e^{ i \frac{1}{2} h \Delta t \sum_{j=1}^{L}  Z_j} U^{(1)}(t) e^{ - i \frac{1}{2} h \Delta t \sum_{j=1}^{L}  Z_j}
\end{eqnarray}
because Z-Pauli rotations commute with the observable and apply only a phase to the initial state.

\subsubsection{2D transverse-field Ising model}
\label{subsubsec:2D}
The quantum quench dynamics of magnetization in the 2D transverse-field Ising model has been studied by means of infinite projected entangled pair state (iPEPS) \cite{Czarnik2019, Dziarmaga2021, Dziarmaga2022} and neural network quantum state \cite{Schmitt2020} calculations. While the iPEPS simulations correspond to dynamics in the thermodynamic limit, neural network simulations were performed on a $10 \times 10$ lattice, which was shown to be sufficiently large to exhibit negligible finite-size effects \cite{Schmitt2020}. In our simulations, we used an $11 \times 11$ square lattice. We set $h=h_c$, where $h_c = 3.04438(2)$ corresponds to the quantum critical point \cite{Blote2002}, and simulate the magnetization up to $t = 0.92$, where we can compare our results to different update schemes used in iPEPS simulations, namely the full update (FU) \cite{Czarnik2019}, neighborhood tensor update (NTU) \cite{Dziarmaga2021}, and gradient tensor update (GTU) \cite{Dziarmaga2022}.

\begin{figure}[!pth]
    \centering
    \includegraphics[width=0.5\textwidth]{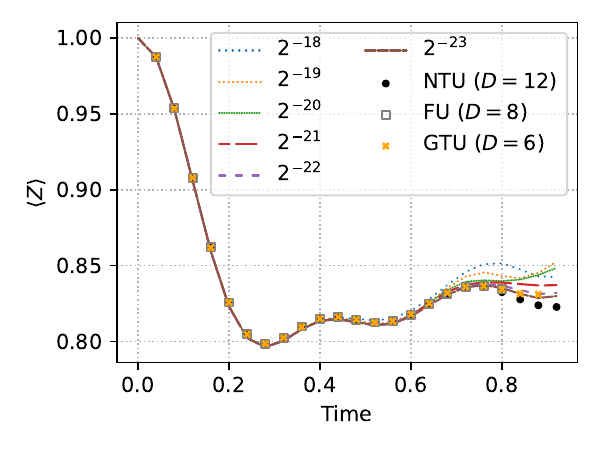}
    \includegraphics[width=0.3125\textwidth]{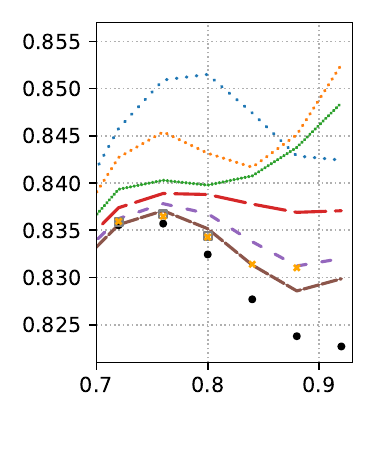}
    \caption{Convergence of xSPD ($M=5$) with respect to threshold and comparison with the iPEPS simulations based on full update (FU) \cite{Czarnik2019}, neighborhood tensor update (NTU) \cite{Dziarmaga2021}, and gradient tensor update (GTU) \cite{Dziarmaga2022} schemes for the $11 \times 11$ 2D square-lattice transverse-field Ising model of Eq.~(\ref{eq:H_tfim_2}) at critical point $h=h_c=3.044382$. The plot in the right panel zooms in on the last part of dynamics ($t>0.7$). All data are plotted at intervals of $\Delta t = 0.04$.}
    \label{fig:2d}
\end{figure}

Figure~\ref{fig:2d} shows the convergence of xSPD with respect to the threshold $\delta$. As expected, the method converges quickly at short times but requires small values of $\delta$ to converge the values at longer times. Our most accurate simulation agrees well with FU and GTU iPEPS results, and shows some deviation from the NTU scheme at the end of the simulation time. 
We note that although the NTU iPEPS calculation corresponds to the largest bond dimension amongst the iPEPS data, the accuracy of truncation is also believed to be less than for the FU iPEPS simulation, thus the relative accuracy of the different reference iPEPS data is unclear.
The disagreement between the two smallest $\delta$ xSPD simulations at the end of the simulation time is only $0.002$ ($0.007$ over the 3 smallest $\delta$s) which provides an estimate of the threshold error. This threshold error is comparable to the estimated Trotter and X-truncation errors (discussed below).

\begin{figure}[!pth]
    \centering
    \includegraphics[width=0.45\textwidth]{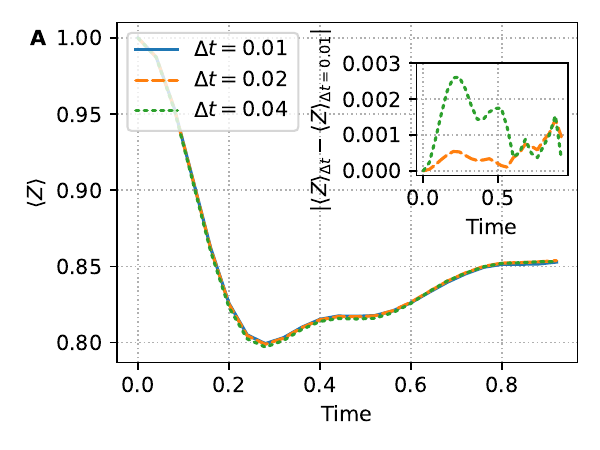}
    \includegraphics[width=0.45\textwidth]{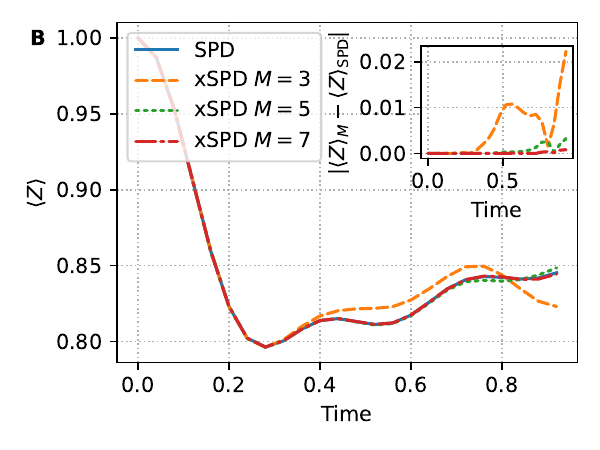}
    \caption{Time step and X-truncation errors for the $11\times 11$ 2D square lattice model at critical point $h=h_c=3.044382$. A: We estimate the time step error by comparing different time steps at a fixed $\delta / \Delta t$ ratio. We find that the Trotter error for $\Delta t = 0.04$ is less than 0.005. B: The error of using xSPD, for truncation with $M=3, 5, 7$. We find that $M=5$ is optimal choice for the total time and accuracy we target in our simulation, with an error below 0.005.}
    \label{fig:2d-checks}
\end{figure}

In this example, we employed a time step of $\Delta t = 0.04$ and set the X-truncation parameter to $M=5$. To validate this choice of parameters, we analyze the associated errors in Fig.~\ref{fig:2d-checks}. Specifically, for the time step (Fig.~\ref{fig:2d-checks}A), we set $\delta / \Delta t = 2^{-19}/0.01  = 1.90734 \times 10^{-4} $ and compute the observable using three different time steps. We estimate that the Trotter error is below $0.003$ within the simulated time of $t=0.92$. Similarly, we perform SPD and xSPD simulations with varying values of $M$, using $\delta=2^{-20}$. We estimate that the error due to $M=5$ X-truncation is about $0.003$. In contrast, employing $M=7$ would lead to almost no error but also limited computational savings, while $M=3$ produces an error that is greater than our convergence target of less than 0.01. Note that due to symmetry, all Pauli operators appear with an even $X$-weight, which is why we only consider odd values of $M$.

Regarding the computational cost, the most accurate simulation at $\delta = 2^{-23}$ generated up to $8.5$ billion Pauli operators, used over 1TB of memory, and took around 36 hours on 16 CPU cores. For comparison, the least accurate simulation shown in Fig.~\ref{fig:2d}, corresponding to $\delta = 2^{-18}$, generated at most 84 million Pauli operators in 32 minutes on 16 CPU cores.

\subsubsection{3D transverse-field Ising model}
\label{subsubsec:3D}
As our final example, we present the quench dynamics of magnetization for spins on a simple cubic lattice with $L=11$ and $n=L^3=1331$ sites. Here we consider two values of $h$, $h=1$ (weak field) and $h=5.15813(6)$ (critical point). 

\begin{figure}[!pth]
    \centering
    \includegraphics[width=0.45\textwidth]{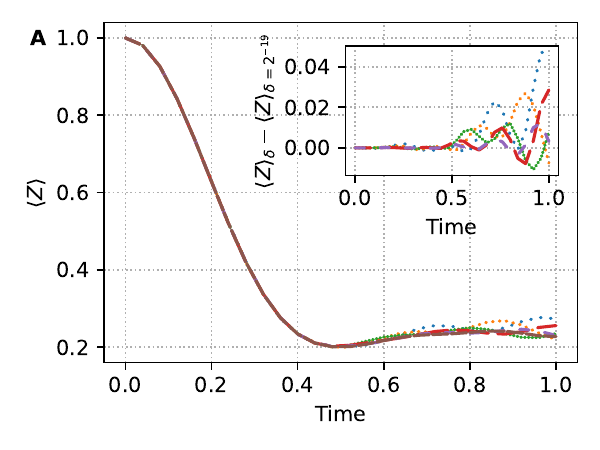}
    \includegraphics[width=0.45\textwidth]{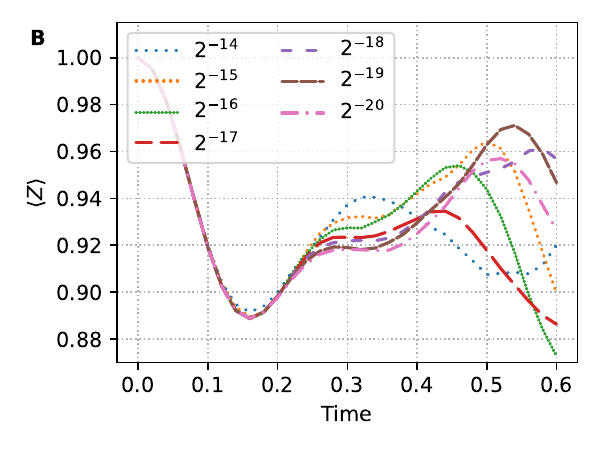}
    \caption{Convergence of xSPD with respect to threshold for the $11 \times 11 \times 11$ 3D tranverse-field Ising model at $h=1$ (panel A, using $\Delta t = 0.04$ and $M=7$) and $h=5.158136$ (panel B, using $\Delta t = 0.02$ and $M=5$). Inset in panel A shows error with respect to the most accurate available simulation ($\delta = 2^{-19}$).}
    \label{fig:3d}
\end{figure}

In the case of $h=1$, using $\Delta t = 0.04$ and $M=7$, we could run the dynamics up to $t=1$ with thresholds as low as $2^{-19}$ (see Fig.~\ref{fig:3d}). The most accurate simulations ($\delta \leq 2^{-17}$) agree to within $\approx 0.02$, which is comparable to the estimated time step (Trotter) and $X$-truncation errors (see Fig.~\ref{fig:3d-checks}A, C in Appendix~\ref{app_sec:3D_params}). Interestingly, even the fastest, least-accurate simulation ($\delta=2^{-14}$) exhibits ``qualitative accuracy'', i.e., recovers the general trend of the most accurate available result. We ascribe this to the fact that sparse Pauli representation can easily reproduce the dynamics dominated by few Pauli operators. However, the method struggles to include small contributions from many Pauli operators generated by the time evolution. For example, while only about $10^6$ Pauli operators are generated at threshold $\delta=2^{-14}$ after 1 time unit, around $7 \times 10^8$ Pauli operators are generated during the same dynamics with a threshold of $\delta=2^{-19}$. Yet, the difference in the observable appears limited to around 0.04.

The system with $h=5.158136$ (Fig.~\ref{fig:3d}B) was propagated up to $t=0.6$ using $\Delta t= 0.02$ and $M=5$. For this choice of parameters, the Trotter and X-truncation errors are estimated to be below 0.005 (see Fig.~\ref{fig:3d-checks}B, D in Appendix~\ref{app_sec:3D_params}). Because of the shorter dynamics we could use a smaller value of the X-truncation parameter $M$ compared to the weak-field case. The results, using thresholds as low as $\delta=2^{-20}$, are converged to below 0.01 for times $t<0.5$, after which our most accurate simulations begin to deviate from each other. This example also illustrates one of the limitations of SPD when it comes to converging results with respect to threshold. Namely, the expectation values are not monotonically converging to the exact result, implying that it is difficult to perform zero-threshold extrapolation, at least far from convergence. Nonetheless, as we approach converged results, such as at smaller thresholds in the 2D square lattice example (Fig.~\ref{fig:2d}), we begin to observe systematic improvement as we reduce the threshold.

In general, these 3D calculations are expected to pose challenges for tensor network techniques, which for a fixed bond dimension show an exponential scaling with the connectivity of the lattice (assuming site tensors with a number of bonds equal to the number of neighbours). Although SPD does not show this exponential scaling, the 3D simulations for a given side-length $L$ are still more costly than the 2D ones, primarily because the number of sites $n=L^3$ is a factor of $L$ larger than in the 2D case.
For these reasons, we cannot simulate as many Pauli operators as in the 2D case. Specifically, our memory budget of about 1.5TB is reached already with less than $10^9$ Pauli operators, an order of magnitude less than in our 2D square lattice simulations. With this number of Pauli operators, the computation with $h=1$ takes around 73 h on 16 CPU cores, about two times longer than our most accurate 2D calculation. Nonetheless, the relative feasibility of these simulations illustrate the kinds of systems that can be (approximately) studied by SPD dynamics that would otherwise be challenging to consider.

\subsubsection{Computational savings from X-truncation}
\label{subsubsec:xspd_savings}
\begin{figure}[!pth]
    \centering
    \includegraphics[width=0.315\textwidth]{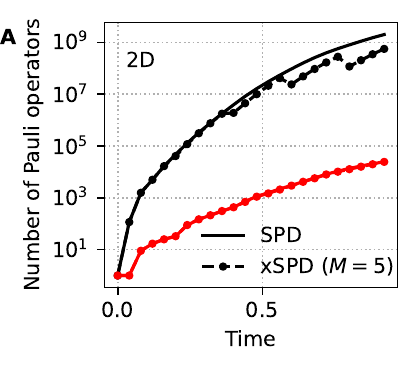}
    \includegraphics[width=0.315\textwidth]{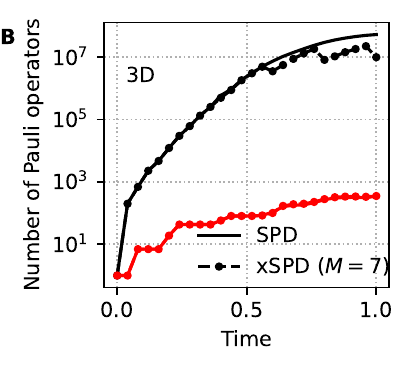}
    \includegraphics[width=0.315\textwidth]{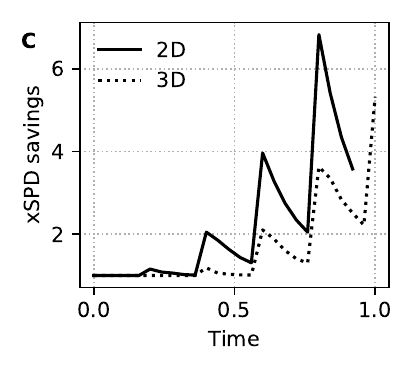}
    \caption{The effect of X-truncation on the number of Pauli operators. A, B: Number of all (black) and Z-type (red) Pauli operators generated with SPD (solid) or xSPD (dashed) for the 2D square lattice (A, $h=3.044382$, threshold $\delta = 2^{-20}$) and 3D cubic lattice (B, $h=1$, threshold $\delta = 2^{-16}$). C: Ratio between the number of Pauli operators in the observable of SPD and xSPD as a function of time.}
    \label{fig:npauli}
\end{figure}

We now analyze the savings due to the X-truncation scheme. Figures~\ref{fig:npauli}A, B show the number of Pauli operators as a function of time for 2D and 3D simulations. While the number of Pauli operators in xSPD is suppressed by the X-truncation, the number of Z-type operators (Pauli operators composed only of $Z$ and identity matrices) is almost the same as in SPD. Since the total cost of the computation is proportional to the number of all Pauli operators $N$, we can use the ratio $N_{\rm SPD}/N_{\rm xSPD}$ to quantify the computational savings (Fig.~\ref{fig:npauli}C). We observe a factor of 6 decrease in the number of Pauli operators in the 2D case, and up to a factor of 5 for the 3D simulation. Such savings allow a factor of 2--4 lower thresholds compared to the original SPD. For example, within a budget of a few days and 1.5 TB of memory, the most accurate SPD calculation we could run for the 2D system would be limited to $\delta=2^{-21}$, which is not converged with respect to threshold at the longest simulated times (see Fig.~\ref{fig:2d}). In contrast, with xSPD we could afford to run the same simulation with $\delta=2^{-23}$.

\section{Conclusion}
\label{sec:conclusion}
To conclude, we have presented numerical simulations of real-time operator evolution with SPD and its modified version, xSPD. We have shown that, in systems for which reference data is available, the performance of these methods is at the level of state-of-the-art tensor network approaches, while the flexibility and simplicity of SPD allows for applications to dynamics problems where tensor network approaches have yet to make an impact, such as in 3D lattices. For the studied systems, we found that we can obtain very accurate results either by converging the observables in the case of short-time, 2D and 3D transverse field Ising model dynamics, or by extrapolating to zero threshold for the long-time diffusion coefficients in 1D chains. Apart from reaching quantitatively converged results for these challenging examples of time-dependent observables, we also found that SPD could serve as a practical method for computing qualitatively accurate results. In this respect, our work motivates further research into using sparse Pauli representations for real-time quantum dynamics.

\begin{acknowledgments}
We thank Jacek Dziarmaga for sharing the iPEPS simulation data presented in Fig.~\ref{fig:2d} and Huanchen Zhai for help with setting up MPO simulations presented in Fig.~\ref{fig:1d-mpo}.
The authors were supported by the US Department of Energy, Office of Science, Office of Advanced Scientific Computing Research and Office of Basic Energy Sciences, Scientific Discovery through Advanced Computing (SciDAC) program under Award Number DE-SC0022088. TB acknowledges financial support from the Swiss National Science Foundation through the Postdoc Mobility Fellowship (grant number P500PN-214214). GKC is a Simons Investigator in Physics. Computations presented here were conducted in the Resnick High Performance Computing Center, a facility supported by Resnick Sustainability Institute at the California Institute of Technology.
\end{acknowledgments}

\section*{Data availability}
The data that support the findings of this article are openly available in a Zenodo repository \cite{zenodo_repo}. The code used to produce the results is available at \url{https://github.com/tbegusic/spd}.

\appendix

\section{Additional details of SPD implementation and working equations}
\label{app_sec:spd}
SPD is implemented as described in Ref.~\cite{Begusic2024}. Briefly, a sum of $N$ Pauli operators is stored in the form of three arrays: an array of $N$ complex coefficients $a$, an array of $N$ integer phases $\varphi$, and a $N \times 2n_{64}$ array of 64-bit unsigned integers that stores two bitstrings $x$ and $z$ for each Pauli operator:
\begin{equation}
    O = \sum_{j=0}^{N-1} a_j (-i)^{\varphi} \prod_{k=0}^{n-1} Z_k^{z_{jk}} X_k^{x_{jk}}. 
\end{equation}
The number of unsigned integers needed to store $n$ bits is $n_{64} = \lceil n \rceil$. Pauli operators are sorted using lexicographic ordering on the bitstrings. In this way, we can find the position $j$ of a given Pauli operator in the sum---or the position at which to insert a new Pauli operator so that the ordering is preserved---in $\mathcal{O}(\log N)$ time. Similarly, deleting Pauli operators preserves the order trivially.

Apart from searching, inserting, and deleting Pauli operators, other key operations on this sparse representation of a sum of Pauli operators include identifying which Pauli operators in the sum anticommute with a given Pauli operator and multiplying the sum of Pauli operators by a Pauli. For the anticommutation of Pauli operators $A$ and $B$, we have
\begin{equation}
    A \text{ anticommutes with } B = z_A \cdot x_B - x_A \cdot z_B.
\end{equation} 
Here the multiplications on the right-hand side correspond to the \textsc{AND} logical operator between bits and additions to the \textsc{XOR} logical operator (addition in $\mathbb{Z}_2$). The product of two Pauli operators $C = A B$ is given by
\begin{eqnarray}
    (z_C, x_C) &=& (z_A + z_B, x_A + x_B), \\
    \varphi_C &=& \varphi_A + \varphi_B + 2 z_A \cdot x_B, \\
    a_C &=& a_A a_B.
\end{eqnarray}

In Sec.~\ref{subsec:1D}, we also introduced an overlap (inner product) between sums of Pauli operators $\text{Tr}[O_1^{\dagger} O_2]/2^n$ represented in the sparse Pauli format described above. Let us assume that $N_1 < N_2$, i.e., that $O_1$ has fewer Pauli operators than $O_2$. Then, we can search for all Pauli operators of $O_1$ in $O_2$ in $\mathcal{O}(N_1 \log N_2)$ time and the overlap is 
\begin{equation}
    \frac{1}{2^n}\text{Tr}[O_1^{\dagger} O_2] = \sum_j (-i)^{\varphi_{2, k[j]} - \varphi_{1, j}} a_{1, j}^{\ast} a_{2, k[j]},
\end{equation}
where the sum runs over Pauli operators in $O_1$ that were found in $O_2$ and $k[j]$ is the index of the $j$-th found Pauli in $O_2$. The expectation over the all-zero state, as needed in Sec.~\ref{subsec:exp_val}, can be computed as
\begin{equation}
    \langle 0^{\otimes n} | O | 0^{\otimes n} \rangle = \sum_j a_j (-i)^{\varphi_j},
\end{equation}
where the sum runs over Z-type Pauli operators (for which all bits in $x_j$ are 0). Finally, the X-weight of an operator is computed as the number of set bits in the corresponding $x$ array.

For convenience, our implementation interfaces to Qiskit \cite{Qiskit} for setting up the calculations. Specifically, it converts Qiskit's SparsePauliOp into our representation described above that is then used in the simulations.

\section{Details of MPO simulations}
\label{app_sec:mpo}
MPO simulations were performed using the time-dependent density matrix renormalization group (TD-DMRG) method, as implemented in the \textsc{Block2} code \cite{Zhai2023}. After constructing the MPO of the observable at time zero, we convert it to an MPS $| O \rrangle$ with a doubled number of sites, i.e., $\langle i_1 i_2 \dots | O | j_1 j_2 \dots \rangle / 2^{n/2} = \llangle i_1 j_1 i_2 j_2 \dots | O \rrangle$. The sites of the MPS are ordered so that the two physical legs of a single site in the MPO appear on neighboring sites in the MPS. The Liouvillian superoperator $L |O\rrangle \equiv [H, O]$ that governs the dynamics of the observable is then represented as an MPO in the extended Hilbert space with twice as many sites. Specifically, each Pauli operator in the Hamiltonian corresponds to a sum of two Pauli operators in the Liouvillian. For single-site Pauli operators $\sigma_i \in \{I, X, Y, Z\}$ at site $i$, we have $\sigma_{2i} - \sigma_{2i+1}$, while the nearest-neighbor interaction terms $\sigma_i \sigma_{i+1}$ correspond to $\sigma_{2i} \sigma_{2i+2} - \sigma_{2i+1} \sigma_{2i+3}$ in the superoperator picture.

The initial observable MPS is propagated with the Liouvillian MPO using the time-step-targeting method \cite{Feiguin2005, Ronca2017} and a time step of $\Delta t = 0.04$. The correlation functions of the form $\text{Tr}[O_1^{\dagger} O_2]/2^n$, used in Eq.~(\ref{eq:C_j_t}), were evaluated as the inner product $\llangle O_1 | O_2 \rrangle$ of the two matrix product states.

\begin{table*}[pth]
    \centering    \caption{\label{tab:runtimes}Runtimes of MPO and SPD simulations for the 1D tilted-field Ising model dynamics shown in Fig.~\ref{fig:1d-mpo}, $L=21$. Calculations were performed on 6 cores of an Intel Xeon Platinum 8352Y processor (2.2 GHz).}
    \begin{ruledtabular}
    \begin{tabular}{ccccc}
    \multicolumn{2}{c}{MPO}
    &&
    \multicolumn{2}{c}{SPD}
    \\\cline{1-2}\cline{4-5}
        $\chi$ & Runtime (minutes) && $\delta$ & Runtime (minutes) \\\hline
        $2^6$ & 14 && $2^{-13}$ & 5 \\
        $2^7$ & 44 && $2^{-14}$ & 21 \\
        $2^8$ & 256 && $2^{-15}$ & 89 \\
        $2^9$ & 1901 && $2^{-16}$ & 453 \\
    \end{tabular}
    \end{ruledtabular}
\end{table*}

\section{Time step and X-truncation parameters in 3D simulations}
\label{app_sec:3D_params}

\begin{figure}[H]
    \centering
    \includegraphics[width=0.45\textwidth]{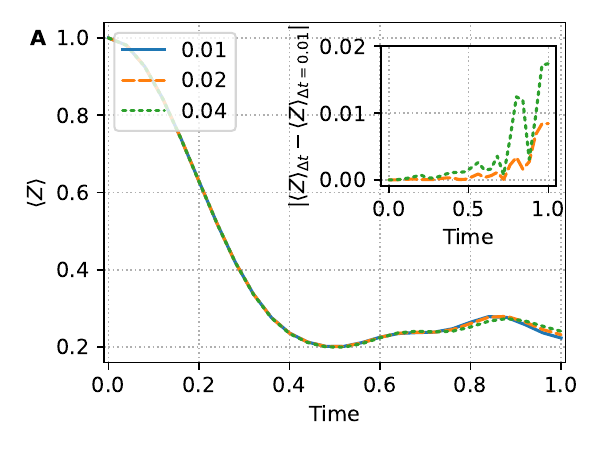}
    \includegraphics[width=0.45\textwidth]{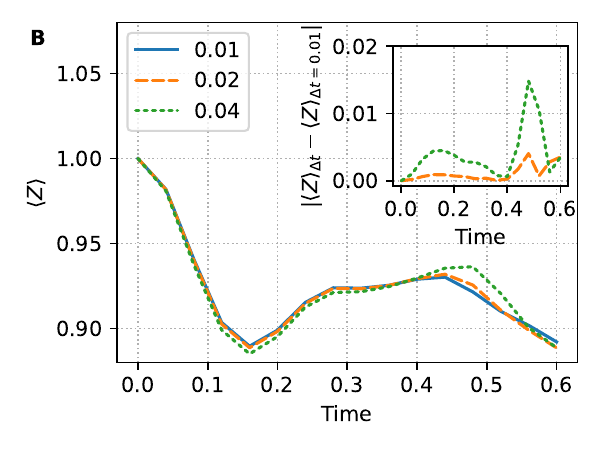}
    \includegraphics[width=0.45\textwidth]{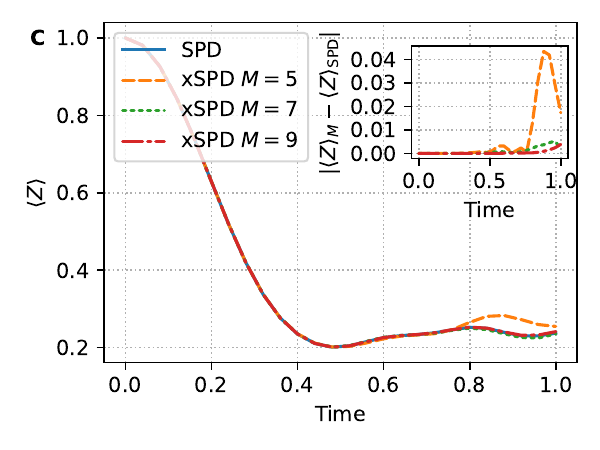}
    \includegraphics[width=0.45\textwidth]{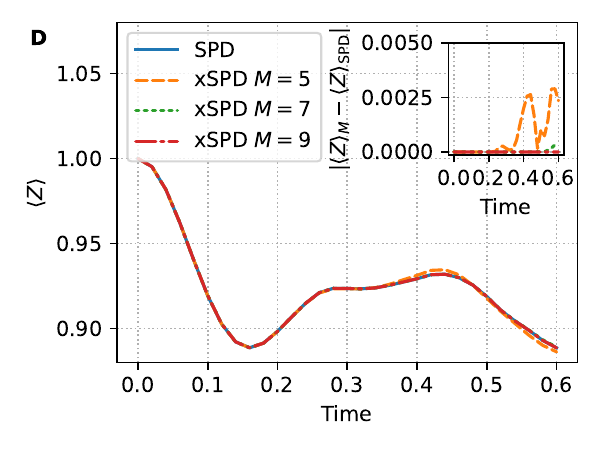}
    \caption{Analogous to Fig.~\ref{fig:2d-checks} but for the 3D transverse-field Ising model with $h=1$ (A, C) and $h=5.158136$ (B, D). For analyzing the time step error (A, B), we set $\delta/\Delta t = 2^{-17}/0.01 \approx 7.63 \times 10^{-4}$ (A) and $\delta/\Delta t = 2^{-18}/0.01 \approx 3.81 \times 10^{-4}$ (B). The estimated Trotter error for $\Delta t = 0.04$ in panel A is less than 0.02, while for $\Delta t = 0.02$ in panel B it is below 0.005. For the xSPD error, we used $\delta = 2^{-16}$ and $\Delta t = 0.04$ for small $h$ (C), and $\delta = 2^{-17}$ and $\Delta t = 0.02$ for critical $h$ (D). With $M=7$, the error in panel C is below 0.006.}
    \label{fig:3d-checks}
\end{figure}

%

\end{document}